\documentclass[%
 aip,
 jap,%
 amsmath,amssymb,amsfont,
 preprint,%
]{revtex4-1}

\usepackage{graphicx}
\usepackage{bm}
\begin{document}

\title{Shape and orientation effects on the ballistic phonon thermal properties of ultra-scaled Si nanowires}%

\author{Abhijeet Paul}
\email{abhijeet.rama@gmail.com}
\author{Mathieu Luisier}
\author{Gerhard Klimeck}

\affiliation{School of Electrical and Computer Engineering, Network for Computational Nanotechnology, Purdue University,\\%
 West Lafayette, Indiana, USA, 47907.}

\date{\today}

\begin{abstract}

The effect of geometrical confinement, atomic position and orientation of Silicon nanowires (SiNWs) on their thermal properties are investigated using the phonon dispersion obtained using a Modified Valence Force Field (MVFF) model. The specific heat ($C_{v}$) and the ballistic thermal conductance ($\kappa^{bal}_{l}$) shows anisotropic variation with changing cross-section shape and size of the SiNWs. The $C_{v}$ increases with decreasing cross-section size for all the wires. The triangular wires show the largest $C_{v}$ due to their highest surface-to-volume ratio. The square wires with [110] orientation show the maximum $\kappa^{bal}_{l}$ since they have the highest number of conducting phonon modes. At the nano-scale a universal scaling law for both $C_{v}$ and $\kappa^{bal}_{l}$ are obtained with respect to the number of atoms in the unit cell. This scaling is independent of the shape, size and orientation of the SiNWs revealing a direct correlation of the lattice thermal properties to the atomistic properties of the nanowires. Thus, engineering the SiNW cross-section shape, size and orientation open up new ways of tuning the thermal properties at the nanometer regime.

%The average power exponents are -0.51 and 0.97 for $C_{v}$ and $\kappa^{bal}_{l}$, respectively. 
%This behavior clearly reveals the atomistic effect on the thermal properties in the ultra-scaled SiNWs.

\end{abstract}


\maketitle %%
\section{Introduction} \label{sec:I}

%\textit{Why SiNWs ?} 

The increasing variety of application of Silicon nanowires (SiNWs) ranging from MOSFETs \cite{Sinwfet} to rechargeable batteries \cite{sinw_battery}, thermoelectric (TE) devices \cite{Hochbaum2008}, bio-sensors \cite{sinw_biosensor}, solar cells \cite{sinw_solar_cells}, etc., requires a good understanding of the lattice thermal properties. Thermal requirements can be very contrasting depending on the type of applications, for eg. MOSFETs will require heat dissipation for better performance whereas thermoelectric devices require low thermal conductivity to maintain a good temperature gradient for higher TE efficiency \cite{jauho_method}.  

Improvement in the process technologies have led to the fabrication of SiNWs with different shapes, sizes and channel orientations \cite{shape_dep_shift_exp,Hochbaum2008,Si_nw_thermal_measure,sinw_synthesis_1}. SiNWs with channel orientations along [100], [110] and [111] are the most commonly manufactured ones.  At the nanometer scale, strong geometrical confinement, atomic positions and increased surface-to-volume ratio (SVR) play significant roles in determining the thermal properties of the SiNWs \cite{jauho_method,mingo_ph,Mingo_kappa}. 

%The phonon spectra of these SiNWs can provide vital information about their thermal properties which can be optimized for maximum benefits 

%\textit{Analysis of thermal properties:} 
%The proper understanding of the thermal properties of SiNWs is crucial to design the of transistors (heat management), thermoelectric (thermal conversion and efficiency) and optical (phonon-photon interaction) devices. 
Thermal transport measurements using techniques like the 3$\omega$ method \cite{sinw_thermal_measure_3w} and thermo-reflectance \cite{sinw_thermoreflectance_meas} have led to a good understanding of the heat flow in large nanowires (diameter $>$ 30nm). However, the physics of heat flow in ultra-scaled SiNWs is still not well understood\cite{jauho_method}. The phonon spectra of SiNWs can provide a theoretical estimate of heat transport through ultra-scaled structured since phonons (lattice vibrations) are responsible for carrying most of the heat in semiconductors \cite{jauho_method,mingo_ph,Mingo_kappa,li_hollow_sinw}.  

In this paper we theoretically explore the effect of (i) cross-section geometry, (ii) cross-section size, and (iii) wire orientation of ultra-scaled SiNWs on the thermal properties such as the ballistic thermal conductance ($\kappa^{bal}_{l}$) and the specific heat ($C_{v}$). Furthermore, analytical expressions for the variation of these physical quantities with size for each cross-section shape and orientation are provided to allow for a compact modeling representation of thermal properties to be used in simulators like Thermal-Spice \cite{Hot_spice} and Themoelectric module simulator \cite{Tspice_tool}.

%\textit{Continuum thermal analysis models:} 
The thermal properties of SiNWs differ considerably from bulk Si \cite{jauho_method,Si_nw_thermal_measure}. The transition from the particle to wave nature of heat transport with structural miniaturization calls for improved heat transport models. The traditional continuum models for thermal conductivity by Callaway \cite{Callaway} and Holland \cite{Holland} are based on the Debye limit for phonons, sound velocity and many other parameters, which render these models quite cumbersome at the nanoscale, as discussed by Mingo et al \cite{mingo_ph}. The dependence of continuum models on a large set of fitting parameters make them unsuitable for predicting the lattice thermal properties with dimensional scaling. This limitation can be overcome by atomistic models which automatically take into account the effects of structural miniaturization like geometrical confinement, orientation effects, cross-sectional shape and surface-to-volume ratio effects \cite{mingo_ph,Mingo_kappa}. 

%The use of phonon dispersion provide very good agreement with experimental thermal measurement in nanostructures and also reveal very interesting atomistic properties in heat transport. 
Theoretical estimates of the thermal conductivity and the specific heat using phonon spectra have been addressed in the literature in the past. Calculations of thermal conductivity using the full phonon spectra of large to medium sized SiNWs along certain specific orientations have been performed \cite{mingo_ph,strain_effect_2}. The influence of surface roughness on nanowires has been studied in Refs.\cite{continuum_model,wire_rough_1,wire_rough_2}. A study of specific heat in $[111]$ SiNWs is reported in Ref. \cite{sinw_111_cv}. Mingo et. al theoretically bench-marked the thermal conductivity of large SiNWs (diameter $\ge$ 34nm) \cite{mingo_ph} and also studied the effect of amorphous coating on the thermal conductivity of SiNWs \cite{Mingo_kappa}. A complete study of orientation effects on the thermal conductivity of SiNW has been provided in Ref. \cite{jauho_method}. The thermal conductivity of Si nano-clusters \cite{si_nano_cluster_1,si_nano_cluster_2} and hollow Si nanowires \cite{li_hollow_sinw,sinw_hollow_therm} has also been studied.

In this work we utilize a semi-empirical phonon model to perform thermal calculations. The phonon dispersion is obtained using a Modified Valence Force Field (MVFF) model \cite{VFF_mod_herman,jce_own_paper}. The continuum models \cite{Holland,Callaway} though computationally simpler and faster, lack the proper physics to extend them to nanostructures due to the use of open fitting parameters \cite{Zi_PRB}.The first principles based models \cite{phon_rev_paper,DFPT_phonon} are limited to extremely small structures (W$<$2nm) with few thousand atoms, and are computationally very expensive. The choice of the MVFF model is guided by the following factors, (i) an adequate amount of physics to understand the thermal properties, and (ii) a lower computational cost thus, allowing the model to handle up to few million atoms for phonon calculations in realistic structures \cite{anharmonic}. At the same time the MVFF model also explains the experimental phonon and elastic properties in zinc-blende materials very well \cite{VFF_mod_herman,VFF_mod_zunger,VFF_kanelis}.

The emphasis of this work is to purely understand the atomistic effects on the thermal properties of ultra-scaled SiNWs such as the effect of the surfaces and atomic coordination number. This understanding is further translated into analytical equations allowing easy usage for other thermal modeling like thermo-electric circuit simulations which require $C_{v}$ and $\kappa_{l}$ as input \cite{Tspice_tool}. Scattering effects are dominating at the nanometer scale in determining the lattice thermal conductivity. However, scattering effects have been neglected in the present study in order to understand the geometric and atomistic effects which will be otherwise convoluted by scattering. Also the ballistic phonon transport can be supported in SiNWs with width smaller than 20nm as pointed in Ref. \cite{casimir_limit}. The scattering effects could be studied as a future work.

%\textit{Paper Organization:} 

The paper is organized as follows. Section~\ref{sec:II} provides a brief description of the MVFF model for the calculation of phonons in SiNWs, the calculation of thermal properties (Sec.~\ref{sec:lat_prop}) and details of the SiNWs used for the study (Sec. \ref{sinw_detail}).
The effect of cross-section shape, size and wire orientation on the specific heat and thermal conductance are discussed in Sec. \ref{sec:Res_3} and Sec. \ref{sec:Res_4}, respectively. This is followed by a discussion on the atomistic effects on the thermal properties in Sec. \ref{sec_discuss}. Conclusions are summarized in Sec. \ref{sec:conc}.

\section{Theory and Approach} 
\label{sec:II}
\subsection{MVFF Phonon model} 
\label{sec_phon_model}

In the MVFF model, the phonon frequencies are calculated from the forces acting on atoms produced by finite displacements of the atoms from their equilibrium positions in a crystal \cite{phon_rev_paper}. First the total potential energy of the solid (U) is estimated from the restoring force(F). In the MVFF model, U is approximated as \cite{jce_own_paper}, 

\begin{eqnarray}
\label{eq:uelastic}
 U & \approx & \frac{1}{2} \sum_{i\in N_{A}} \Bigg[ \sum_{j\in nn(i)} U^{ij}_{bs} + \sum^{j \neq k}_{j,k \in nn(i)} \big( U^{jik}_{bb} \nonumber \\
            & &  + U^{jik}_{bs-bs} + U^{jik}_{bs-bb} \big) + \sum^{j \neq k \neq l}_{j,k,l \in COP_{i}} U^{jikl}_{bb-bb} \Bigg],
\end{eqnarray}

where $N_{A}$, $nn(i)$, and $COP_{i}$ represent the total number of atoms in one unit cell, number of nearest neighbors for atom `i', and the coplanar atom groups for atom `i', respectively. The first two terms $U^{ij}_{bs}$ and $U^{jik}_{bb}$ represent the elastic energy obtained from bond stretching and bending between atoms connected to each other \cite{Keating_VFF}. The terms $U^{jik}_{bs-bs}$, $U^{jik}_{bs-bb}$, and $U^{jikl}_{bb-bb}$ represent the cross bond stretching \cite{VFF_mod_herman}, the cross bond bending-stretching \cite{VFF_mod_zunger}, and the coplanar bond bending \cite{VFF_mod_herman} interactions, respectively. The detailed procedure for obtaining the phonon spectra in bulk Si and NWs are outlined in Ref.\cite{jce_own_paper,iwce_own_paper}. 

%In the following sections the method to extract the acoustic and optical phonon shifts as well as the thermal properties and the types of SiNWs used in this study are discussed. 

%Also the shape effect on the thermal properties of SiNWs are discussed.
% the procedure to obtain the phonon shifts in these nanowires. In the following section we discuss the methods to obtain the phonon shifts for SiNWs once the phonon spectra is obtained using the MVFF model.

\subsection{Lattice thermal properties}
\label{sec:lat_prop}

The complete phonon dispersion provides information about the thermal properties of nanostructures \cite{mingo_ph,Mingo_kappa,jauho_method}. The constant volume temperature (T) dependent specific heat ($C_{v}(T)$) can be evaluated using the following relation \cite{Fph_1},

\begin{eqnarray}
\label{cv_eqn}
	C_{v}(T) & = &  \\ \nonumber 
		& & (\frac{k_{B}}{m_{uc}})\cdot\sum_{n,q} \Big[ \frac{ (\frac{\hbar \omega_{n,q}}{k_{B}T})\cdot \exp(\frac{-\hbar \omega_{n,q}}{k_{B}T})}{[1-\exp(\frac{-\hbar \omega_{n,q}}{k_{B}T})]^{2}} \Big] \; [J/kg K],
\end{eqnarray}
where $k_{B}$, $\hbar$ and $m_{uc}$ are the Boltzmann's constant, reduced Planck's constant, and the mass of the SiNW unit cell in kg, respectively. The quantity $\omega_{n,q}$ is the phonon frequency associated with the branch `n' and crystal momentum vector `q'. 

For a semiconductor slab/wire with a small temperature difference $\Delta T$ between its two extremities, the thermal conductance ($\kappa^{bal}_{l}$) is obtained using Landauer's method\cite{Land} as \cite{mingo_ph,Mingo_kappa,jauho_method},  

\begin{eqnarray}
\label{ktherm_eq}
\kappa^{bal}_{l}(T)& = & \hbar \int_{0}^{\omega_{max}}T(\omega)\cdot M(\omega)\cdot\omega\cdot \\ \nonumber 
	 & &\frac{\partial}{\partial T} \Big[ (\exp(\frac{\hbar \omega}{k_{B}T})-1)^{-1}\Big]\cdot d\omega \;\;[W/K],
\end{eqnarray}

The term $M(\omega)$ is the number of phonon modes at frequency $\omega$ and $T(\omega)$ is the transmission for each mode. For ballistic conductance each mode transmits with a probability of 1 while for coherent scattering dominated conductance the transmission value is less than 1.

\subsection{Si Nanowire details}
\label{sinw_detail}

\begin{figure}[!htb]
	\centering
		\includegraphics[width=3.1in,height=3.1in]{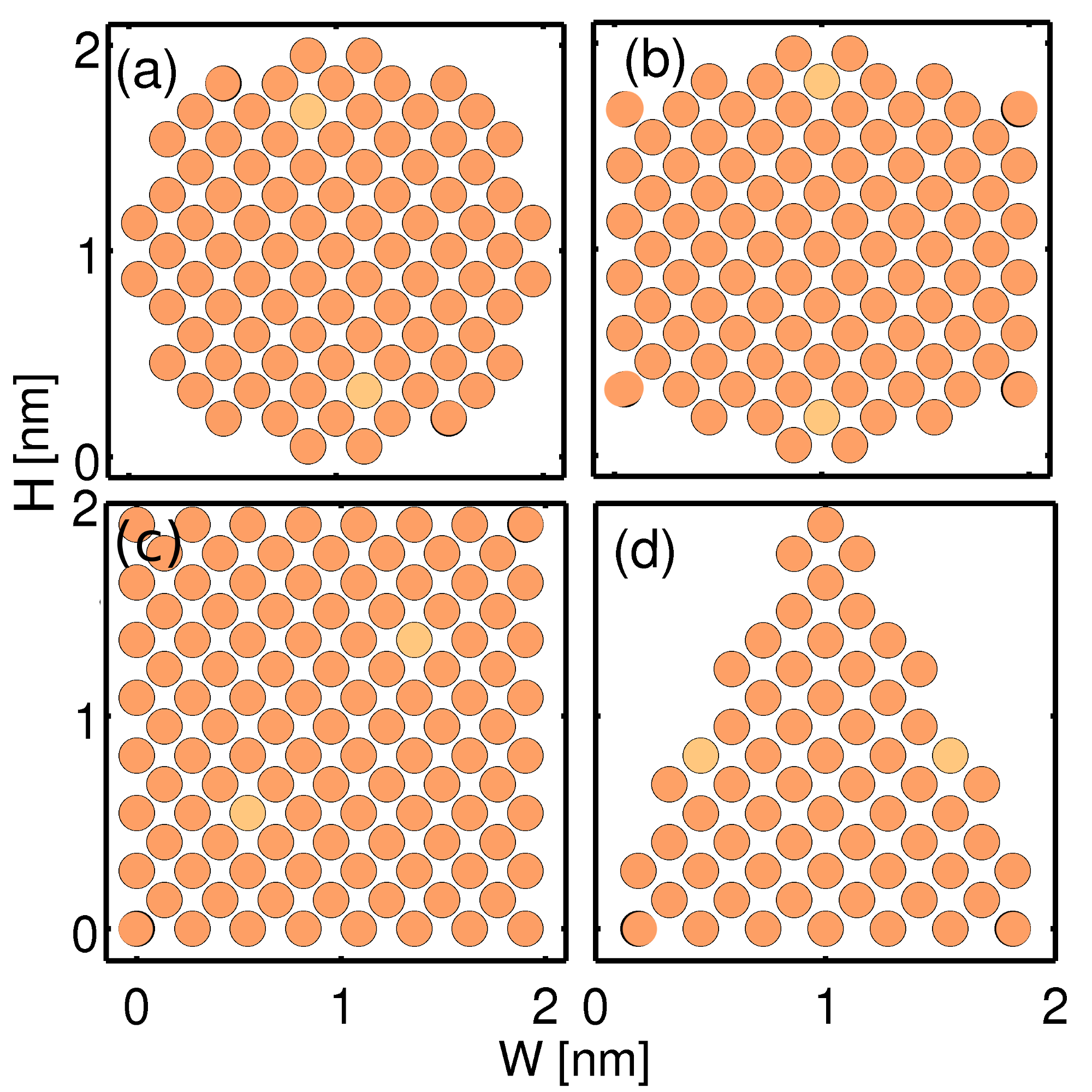}
	\caption{Projected unit cell structures of free-standing $[$100$]$ oriented silicon nanowires with (a) Circular, (b) Hexagonal, (c) Square, and (d) Triangular cross-section shapes. Width and height of the cross-section are defined using a single width variable W (width = height). These structures are at W = 2nm.}
	\label{fig:SiNW_unitcell}
\end{figure}

\begin{figure}[htb!]
	\centering
		\includegraphics[width=3.2in,height=2.0in]{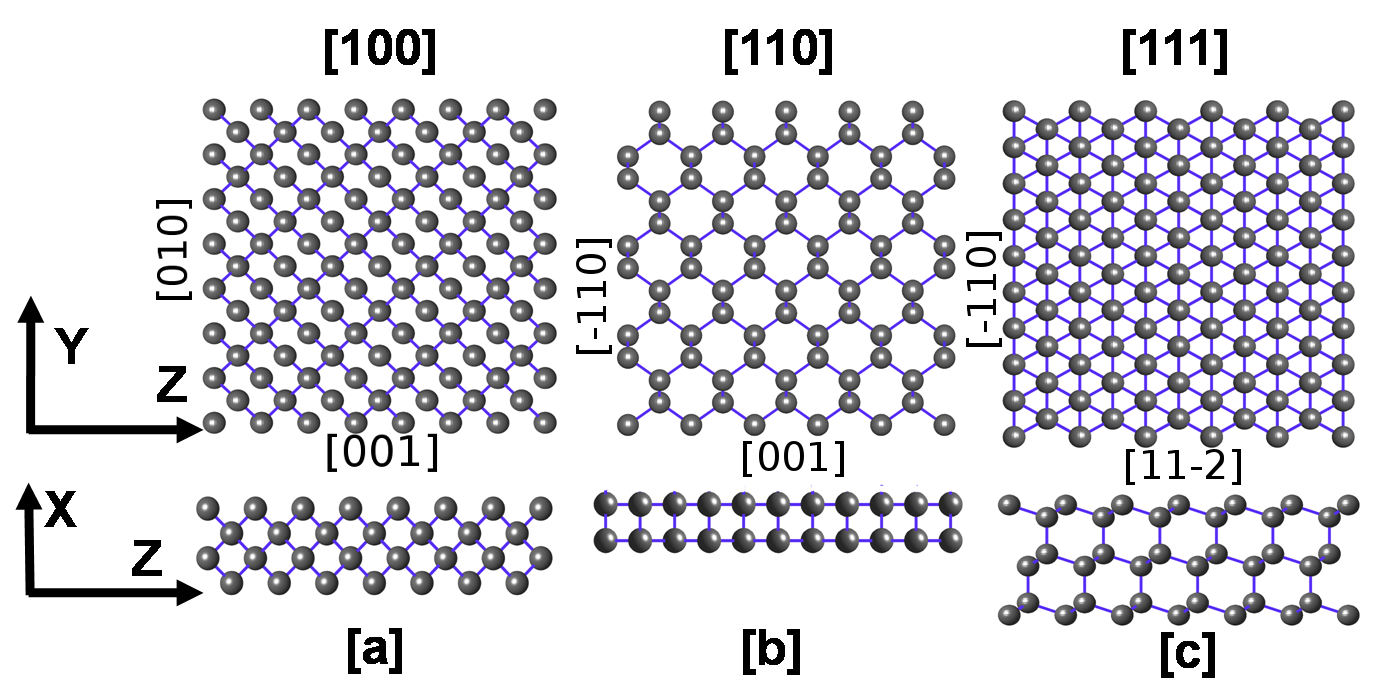}
	\caption{Projected unit cells of square free-standing SiNWs with (a) [100], (b) [110], and (c) [111] wire axis orientation. The width and the height of the SiNWs are defined using W. Here, W = 2nm.}
	\label{fig:SiNW_unitcell_or}
\end{figure}

In this study, four types of cross-section shapes for $[$100$]$ SiNWs have been considered namely, (a) circular, (b) hexagonal, (c) square and (d) triangular (Fig.~\ref{fig:SiNW_unitcell}). Square SiNWs with [110] and [111] channel orientations have been studied too (Fig. \ref{fig:SiNW_unitcell_or}). The feature size is determined by the width parameter W. The value of W is varied from 2 to 6 nm. The confinement direction of the SiNWs are along Y and Z. The heat transport direction is along X. The surface atoms are allowed to vibrate freely without any passivating species. These wires are assumed to have a tetrahedral geometry. It has been shown that wires with diameter below 2nm tend to lose the tetrahedral structure \cite{sinw_stability,Amrit} due to surface pressure and internal strain.

%The details of these calculations are provided in   
 % Similarly the optical phonons also get confined in nanostructures which is given by \cite{hepplestone_sinw_1,jusserand}, 
%where $\omega^{opt}_{bulk}$ is the bulk optical phonon frequency at the BZ center. We utilize Eq. (\ref{ac_phonon_shift}) and (\ref{opt_phonon_shift}) for the calculation of phonon shifts in SiNWs which is discussed next.

\section{Results and Discussion} 
\label{sec:Res}

In this section the results on the effect of cross-section shape, size and orientation on the thermal properties of SiNWs are presented and discussed.  All the thermal quantities are calculated at 300K. However, the analysis holds for any temperature (T) where the anharmonic phonon effects are small. For T $<\;T_{Debye}$ the anharmonic interactions are quite small \cite{li_hollow_sinw}. For bulk Si, $T_{Debye}$ $\sim$725K \cite{shape_depend_theory} and it further increases for SiNWs \cite{sinw_stability}.

\subsection{Specific heat in SiNWs}
\label{sec:Res_3}

%In this section we discuss the impact of wire cross-section shape on the thermal properties of the $[$100$]$ SiNWs. Specific heat ($C_{v}$)  and ballistic lattice thermal conductance ($\kappa^{bal}_{l}$) are the two quantities being discussed here.  

\begin{figure}[!htb]
	\centering
		\includegraphics[width=3.3in,height=1.7in]{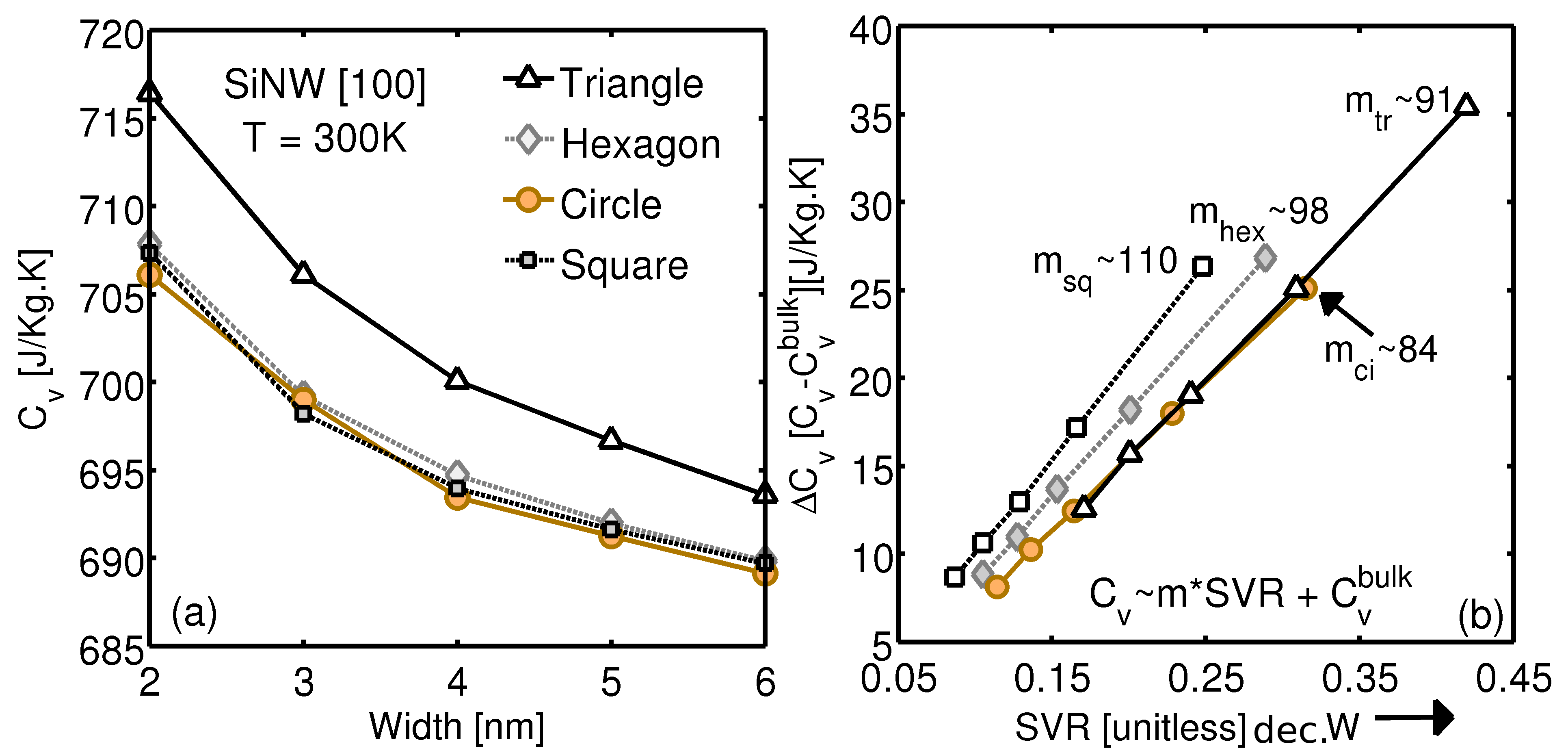}
	\caption{(a) Dependence of the specific heat ($C_{v}$) on cross-section shape and size in [100] SiNWs. (b) Variation in $\Delta C_{v}$ ( = $C_{v}-C^{bulk}_{v}$ ) with SVR for all the [100] SiNW shapes. $C^{bulk}_{v}$ for each shape is given in Table. \ref{table_2}. }	
	\label{fig:cv_shape}
\end{figure}

\textit{Influence of the shape and size on $C_{v}$:} The $C_{v}$ of SiNWs increases with decreasing cross-section size in all the wire shapes (Fig. \ref{fig:cv_shape}a). The size dependence of $C_{v}$ can be approximated by the following relation \cite{sinw_111_cv},

\begin{equation}
	\label{eq_cv_w}
	C_{v}(W) = C^{bulk}_{v} + \frac{A}{W},
\end{equation}
where, A is a fitting parameter extracted from the linear fit of the numerical simulations. Table \ref{table_2} shows the value of $C^{bulk}_{v}$ and A for each geometry. As W $\rightarrow \infty$ (increasing cross-section size), the $C_{v}$ of all the SiNWs converges to a fixed value of $\sim$681 J/kg.K which is reasonably close to the experimental $C_{v}$ value for bulk Si ($\sim$682 J/Kg.K as provided in Refs. \cite{madelung,sinw_111_cv}). The triangular wires show the maximum $C_{v}$ for all the W value, whereas the other shapes show similar $C_{v}$ values at any given cross-section size (Fig. \ref{fig:cv_shape}a).

The plot of $\Delta C_{v}$ (=$C^{wire}_{v}-C^{bulk}_{v}$) vs. SVR (SVR = Total surface Atoms/Total atoms in unit cell) shows a linear behavior (Fig. \ref{fig:cv_shape}b), which can be represented as,

\begin{equation}
	\label{eq_cv_svr}
	C^{wire}_{v} \approx m_{c} \times \text{SVR} + C^{bulk}_{v},
\end{equation}
where $m_{c}$ describes the additional contribution to the $C_{v}$ of the SiNWs with increasing surface-to-volume ratio. The value of $m_{c}$ is positive for all the wire shapes (Fig. \ref{fig:cv_shape}b) which corroborates the fact that specific heat increases with increasing surface area \cite{sinw_111_cv}. Different coordination number of surface atoms for the various cross-section shapes result in different $m_{c}$ values which depict the atomistic effect on the $C_{v}$ value in ultra-scaled SiNWs.

%As the wire size increases the SVR $\rightarrow$ 0 and $C^{wire}_{v}$ converges towards the bulk value.    

The $C_{v}$ increase with decreasing W can be attributed to two phenomena, (i) phonon confinement due to small cross-section size and (ii) an increased surface-to-volume ratio (SVR) in smaller wires \cite{iwce_own_paper,sinw_111_cv}. With increasing geometrical confinement (smaller cross-section size) the phonon bands are more separated in energy \cite{iwce_own_paper} which makes only the few lower energy bands active at a given temperature (see Eq. \ref{cv_eqn}). Thus, more energy is needed to raise the temperature of the smaller wires.

\begin{figure}[htb!]
	\centering
		\includegraphics[width=3.3in,height=1.7in]{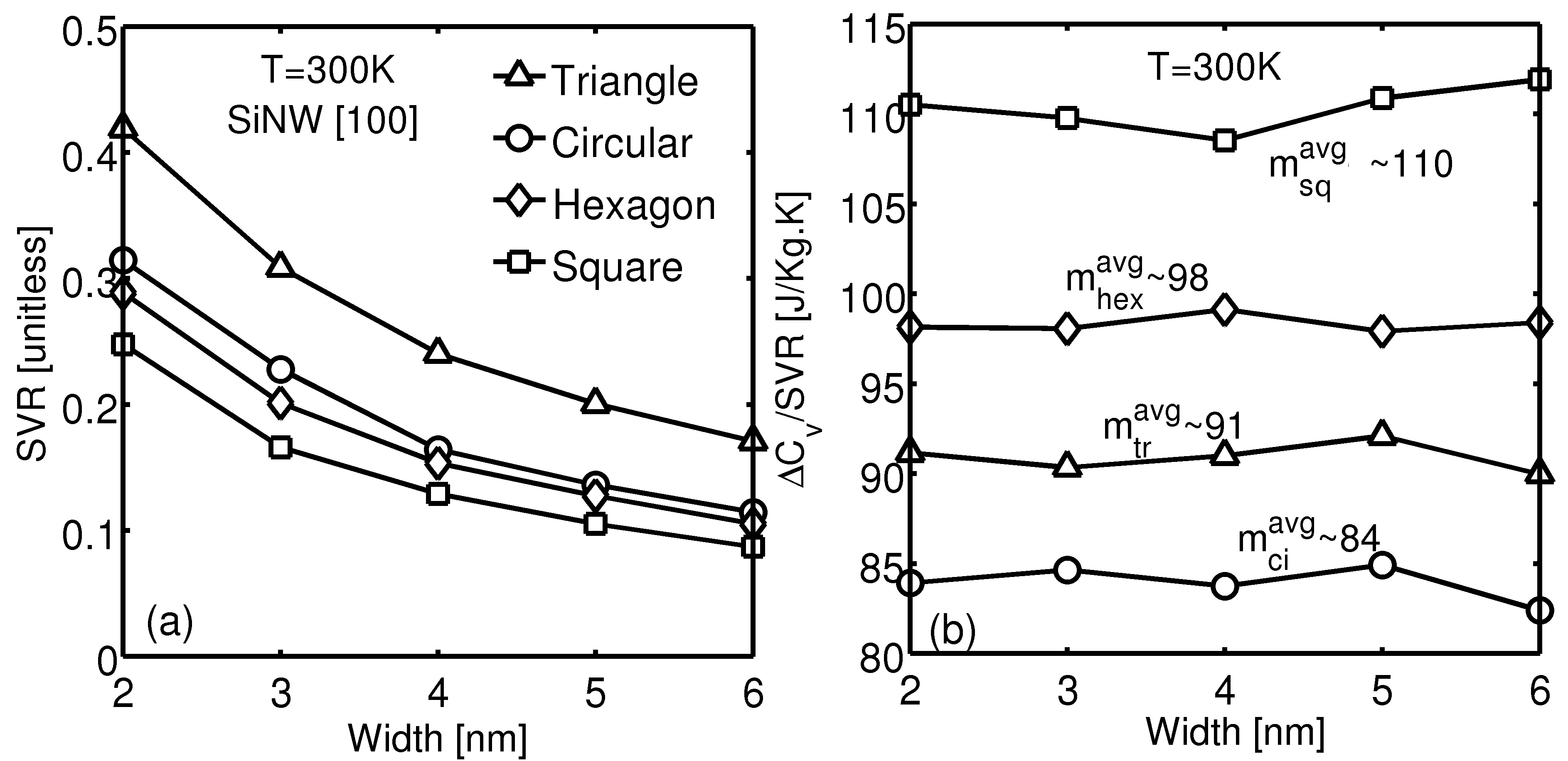}
	\caption{(a) Surface-to-volume ratio (SVR) for different cross-section shape and size [100] SiNWs. (b) Incremental contribution to the specific heat with SVR for different cross-section shape [100] SiNWs.}
	\label{fig:cv_shape_svr}
\end{figure}

The shape dependence of the $C_{v}$ can be understood from Eq. \ref{eq_cv_svr}. The variation of (i) SVR, and (ii) $m_{c}$ with W for different shapes decide the eventual $C_{v}$ order. Figure \ref{fig:cv_shape_svr}(a) shows that triangular wires have the maximum SVR while the other shapes have similar SVR at a fixed W. 
%SVR is calculated as the ratio of the surface atoms to the total number of atoms in the unitcell of the SiNW. 
The increasing SVR results in a higher phonon density of states (DOS) associated with the wire surface, which further enhances the specific heat with decreasing wire cross-section \cite{sinw_111_cv}. The square wires provide the largest surface contribution to $C_{v}$ as depicted by the variation in $m_{c}$ ($\Delta C_{v}$/SVR) (Fig. \ref{fig:cv_shape_svr}b). An optimal value of SVR and $m_{c}$ in SiNWs will maximize the $C_{v}$. The $SVR\;\times\;m_{c}$ value has the following order, triangle ($\sim$36) $>$ hexagonal ($\sim$29) $>$ square ($\sim$27) $>$ circular ($\sim$26). Thus, triangular wires depict the highest $C_{v}$ due to the highest SVR. However, the trends for SVR ($SVR_{sq}\;<\;SVR_{hex}\;<\;SVR_{ci}$) and $m_{c}$ ($m^{sq}_{c}\;>m^{hex}_{c}\;>m^{ci}_{c}$) are opposite for the other shapes, hence resulting in almost similar $C_{v}$ values. The $C_{v}$ values have the following shape order in [100] SiNWs: \textit{triangular $>$ hexagonal $\approx$ square $\approx$ circular.}

\begin{figure}[!htb]
	\centering
		\includegraphics[width=3.3in,height=1.7in]{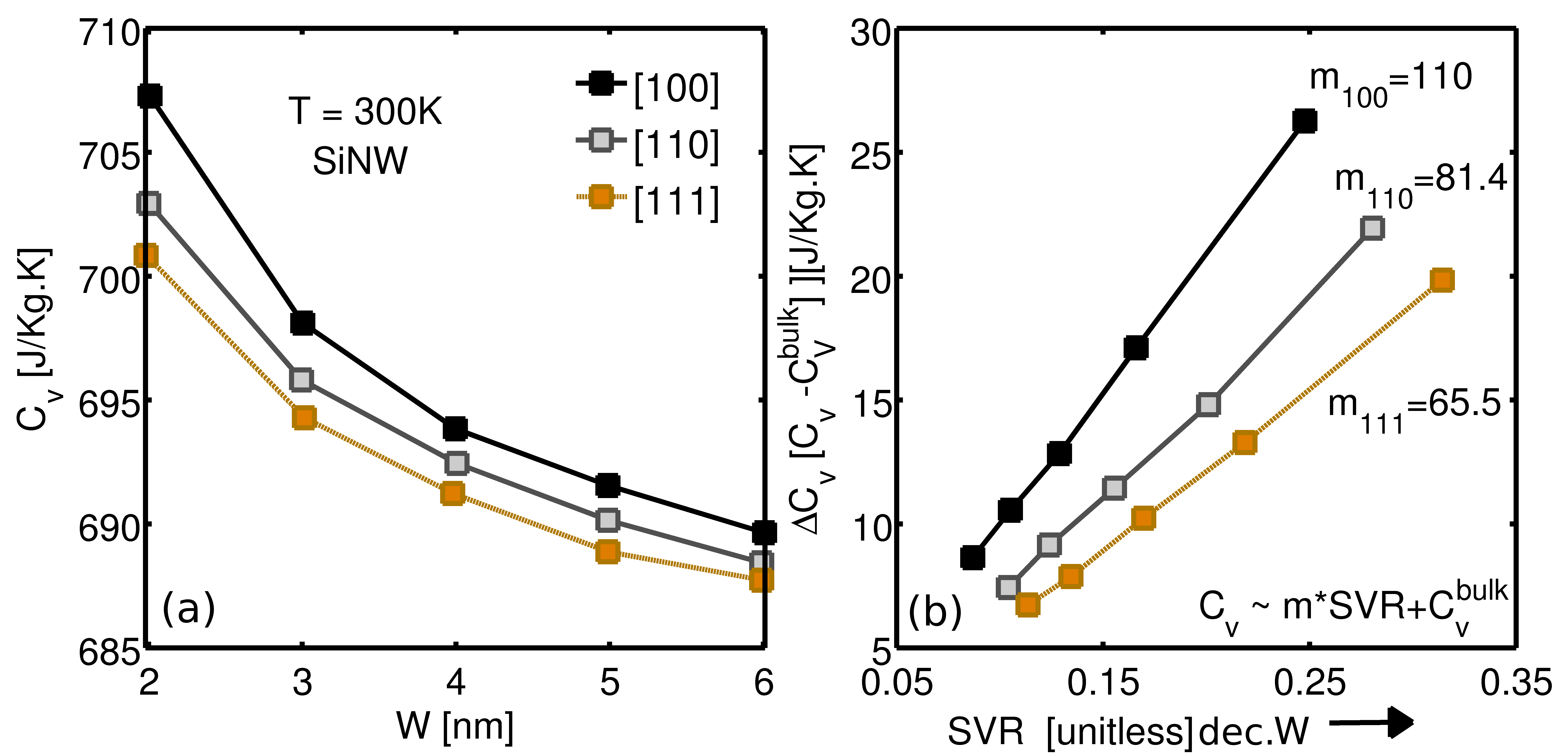}
	\caption{(a) Dependence of the specific heat ($C_{v}$) on the orientation of square SiNWs. (b) Variation in $\Delta C_{v}$ ( = $C_{v}-C^{bulk}_{v}$ ) with SVR for all the SiNW orientations. $C^{bulk}_{v}$ are taken from Table. \ref{table_2_1}.}
	\label{fig:cv_orient}
\end{figure}

\begin{figure}[!htb]
	\centering
		\includegraphics[width=3.3in,height=1.7in]{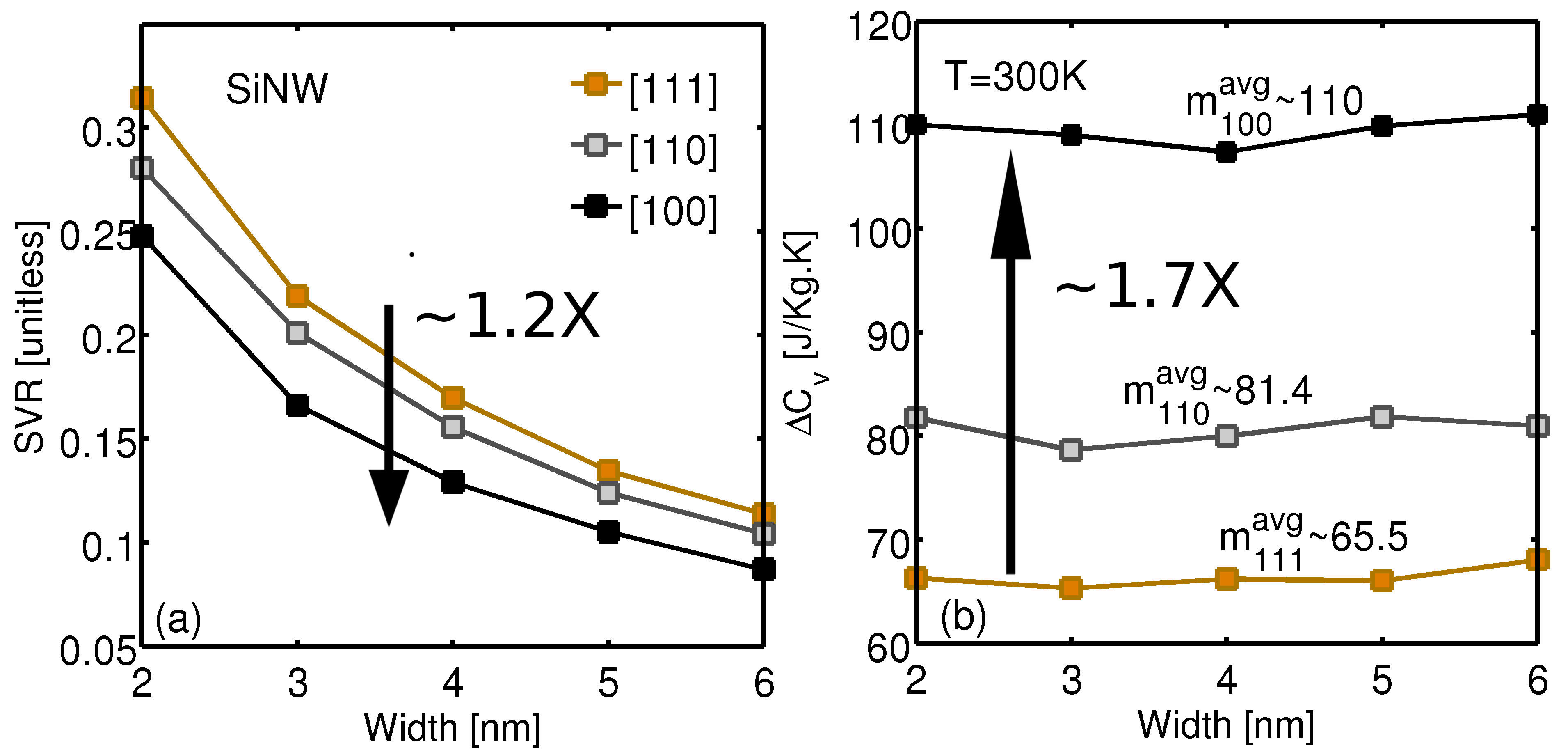}
	\caption{(a) Surface-to-volume ratio (SVR) for different wire orientations of square SiNWs. (b) Incremental contribution to the specific heat with SVR for different orientations of SiNWs.}
	\label{fig:cv_orient_svr}
\end{figure}

\textit{Influence of orientation on $C_{v}$:} The specific heat is also a function of the SiNW orientation (Fig. \ref{fig:cv_orient}a). The $C_{v}$ varies inversely with W, similar to the trend extracted from different shapes (Eq. \ref{eq_cv_w}). The width parameters (Eq. \ref{eq_cv_w}) for the variation of $C_{v}$ with orientation are provided in Table \ref{table_2_1}. The $\Delta C_{v}$ value again shows a linear variation with SVR for different wire orientations (Fig. \ref{fig:cv_orient}b). The $C_{v}$ has the following trend with orientation for different cross-section sizes, $C^{100}_{v}\;\;>C^{110}_{v}\;\;>C^{111}_{v}$.  This trend can be explained again by looking at the impact of (i) SVR and (ii) $m_{c}$ ( Eq. \ref{eq_cv_svr}) on the overall $C_{v}$ value. The SVR shows the following order with W, $\text{SVR}^{111}\;\;>\text{SVR}^{110}\;\;>\text{SVR}^{100}$ (inc. of $\sim$1.2$\times$ from [100] to [111]) as illustrated in Fig. \ref{fig:cv_orient_svr}a. However, $m_{c}$ shows the following  order with W, $m^{100}_{c}\;\;>m^{110}_{c}\;\;>m^{111}_{c}$ (dec. of $\sim$1.7$\times$ from [100] to [111])  (Fig. \ref{fig:cv_orient_svr}b). \textit {The larger surface to volume ratio plays the main role in deciding the $C_{v}$ trend for SiNWs with different orientations.}

\subsection{Ballistic thermal conductance of SiNWs}
\label{sec:Res_4}
The thermal conductance indicates how a structure can carry heat. For high thermoelectric efficiency (ZT) a small thermal conductance is needed \cite{shape_dep_ZT_GeNW,Mingo_kappa,mingo_ph} whereas for CMOS transistors a high $\kappa_{l}$ is preferred  to evacuate the heat. The cross-section shape, size and wire orientation of SiNWs can be used to tune their the thermal conductance.

\begin{figure}[!htb]
	\centering
		\includegraphics[width=3.3in,height=1.7in]{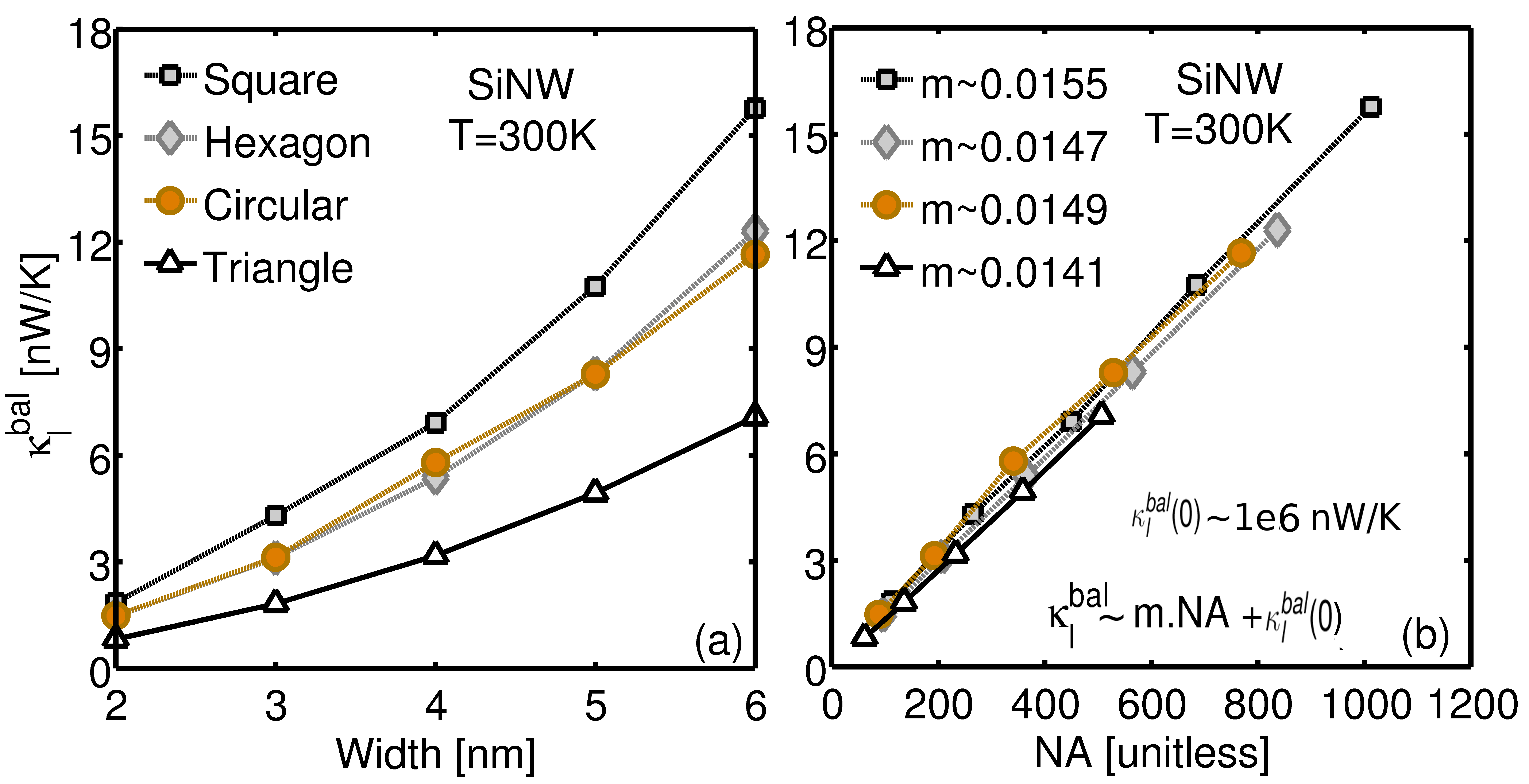}
	\caption{(a) Effect of cross-section shape and size on the ballistic thermal conductance of $[$100$]$ SiNWs. (b) Variation in $\kappa^{bal}_{l}$ with the total number of atoms per unit cell (NA) for different cross-section shapes.}
	\label{fig:kthermal_shape}
\end{figure}

\textit{Influence of the shape and size on $\kappa^{bal}_{l}$:} The ballistic thermal conductance ($\kappa^{bal}_{l}$) for 4 different shapes (Fig. \ref{fig:SiNW_unitcell}) of [100] SiNWs are calculated as shown in Fig. \ref{fig:kthermal_shape}a. The $\kappa^{bal}_{l}$ can be fitted according to the following size (W) relation,

\begin{equation}
	\label{eq_kbal_w}
	\kappa^{bal}_{l}(W) =  \kappa_{0}\Big(\frac{W}{a_{0}}\Big)^{d},
\end{equation}
where $a_{0}$ is the silicon lattice constant (0.5431 nm), d is a power exponent, and $\kappa_{0}$ is a constant of proportionality. The values of `d' and $\kappa_{0}$ for different wires shapes are provided in Table \ref{table_3}. The value of d varies between 1.92 and 2.011 which implies that $\kappa^{bal}_{l}$ has a similar size dependence for all the wire shapes. However, the pre-factor value ($\kappa_{0}$) reflects the shape dependence. This pre-factor has the same ordering as the thermal conductance ordering (Fig. \ref{fig:kthermal_shape}a and Table \ref{table_3}).   

%An interesting property obtained from the calculations is that 
The ballisitic thermal conductance exhibits a linear behavior with the number of atoms per unit cell (NA) as depicted in Fig. \ref{fig:kthermal_shape}b. This linear relation can be approximated by the following equation,
\begin{equation}
	\label{eq_kbal_na}
	\kappa^{bal}_{l} \approx  m_{k}\times NA + \kappa^{bal}_{l}(0),
\end{equation}
where the slope $m_{k}$ represents the average contribution from each atom in the unit cell to $\kappa^{bal}_{l}$ and $\kappa^{bal}_{l}(0)$ is the thermal conductance at NA = 0. The value of $\kappa^{bal}_{l}(0)$ is zero within numerical error ($\kappa^{bal}_{l}(0)\approx1e-7$) which is expected for NA = 0. The value of $m_{k}$ takes into account the surface, shape and atomic effects since the calculation procedure involves the complete phonon dispersion. This relation shows a direct correlation of the atomistic effects to the ballistic thermal conductance.

The size dependence can be explained by the fact that the larger wires have (i) more phonon sub-bands resulting in higher number of modes ($M(\omega)$ in Eq. \ref{ktherm_eq}) and (ii) a higher acoustic sound velocity which is responsible for a larger heat conduction in these SiNWs \cite{iwce_own_paper}.

\begin{figure}[!t]
	\centering
		\includegraphics[width=3.4in,height=1.72in]{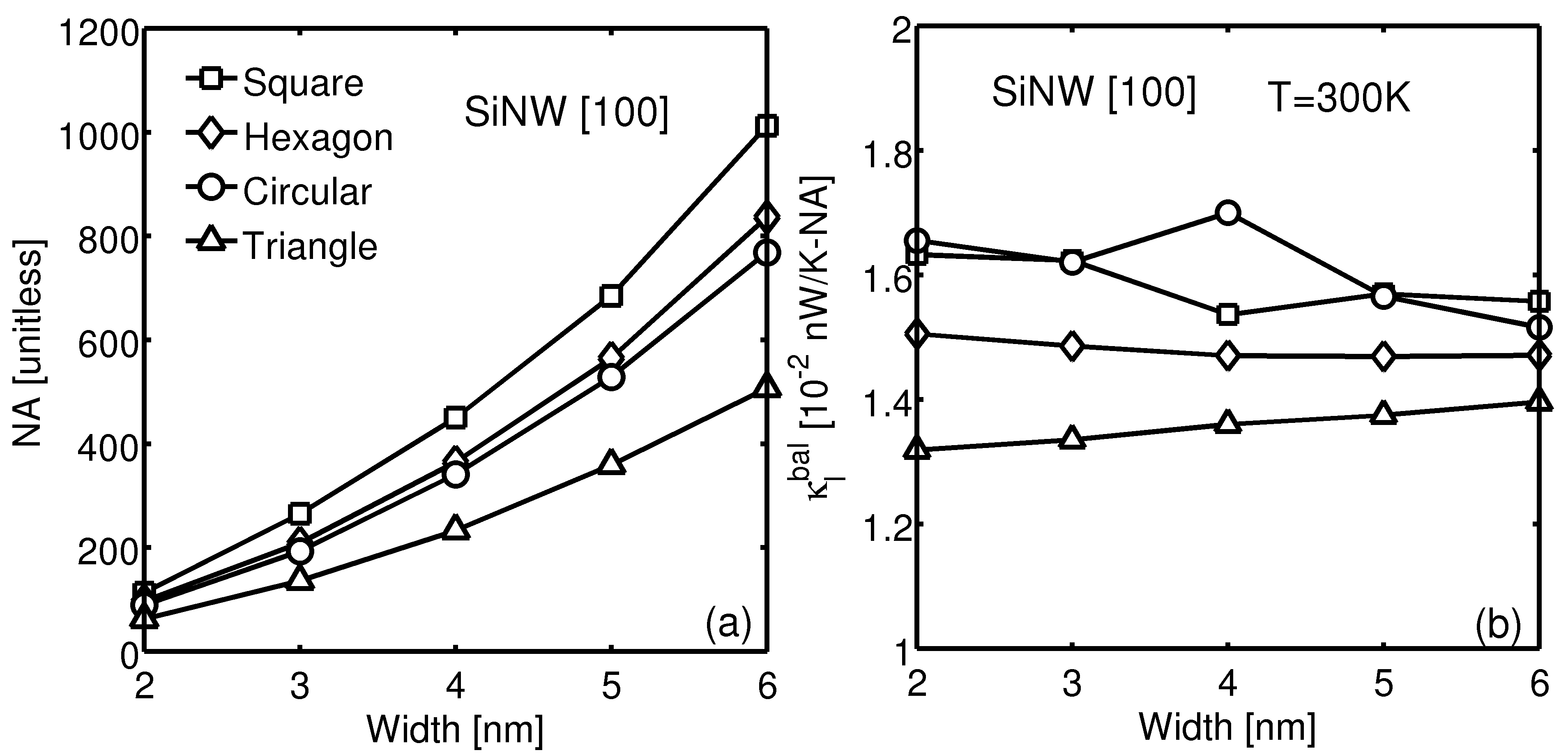}
	\caption{(a) The number of atoms (NA) in [100] SiNW unit cell for different cross-section size and shapes. (b) Contribution to $\kappa_{l}$ per atom for different cross-section shapes.}
	\label{fig:kthermal_shape_NA}
\end{figure}

The shape dependence can be explained as an interplay of two effects, (i) the total number of atoms (NA) present in the unit cell of SiNWs, and (ii) the average contribution of every atom towards $\kappa^{bal}_{l}$. For a fixed cross-section size W, the NA ordering is $NA_{sq}\;\;>NA_{ci}\;\;>NA_{hex}\;\;>NA_{Teri}$ (Fig. \ref{fig:kthermal_shape_NA}a). The total number of phonon branches are 3$\times$NA, due to the three degrees of freedom associated with each atom \cite{jce_own_paper}.  The number of phonon modes ($M(\omega)$) is directly proportional to the phonon energy sub-bands. The contribution per atom to the thermal conductance ($m_{k}$) stays almost constant with the wire cross-section size for a given shape (Fig. \ref{fig:kthermal_shape_NA}b). \textit{ Since, the values of $m_{k}$ are quite similar for all the cross-section shapes , the ordering of NA with shape governs the dependence of $\kappa^{bal}_{l}$ on the cross-section shape.}

\begin{figure}[!htb]
	\centering
		\includegraphics[width=3.4in,height=1.72in]{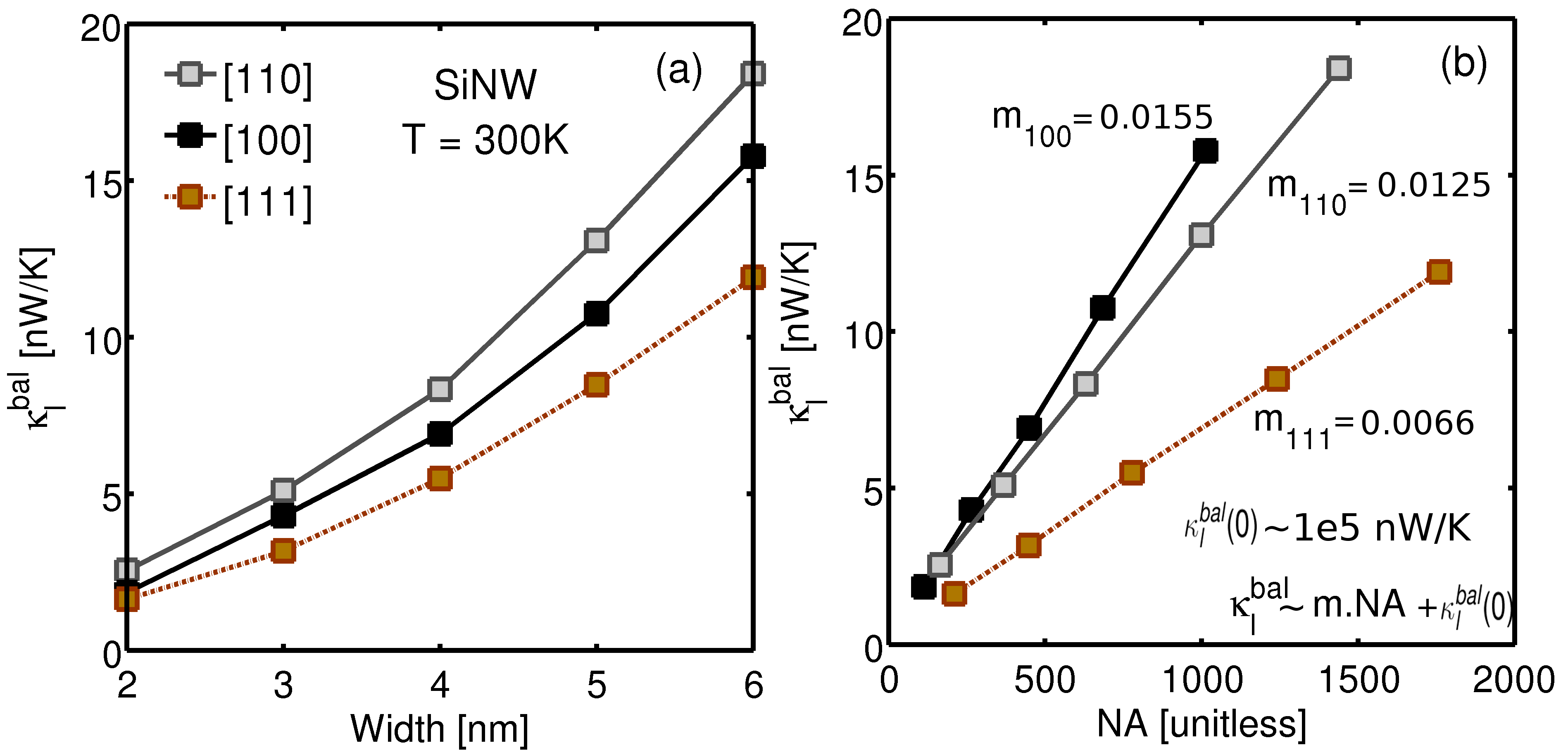}
	\caption{(a) Effect of size on the ballistic thermal conductance in SiNWs with different orientations. (b) Variation in $\kappa^{bal}_{l}$ with the total number of atoms in the unit cell (NA) for different wire orientations.}
	\label{fig:kthermal_orient}
\end{figure}

\begin{figure}[!htb]
	\centering
		\includegraphics[width=3.4in,height=1.72in]{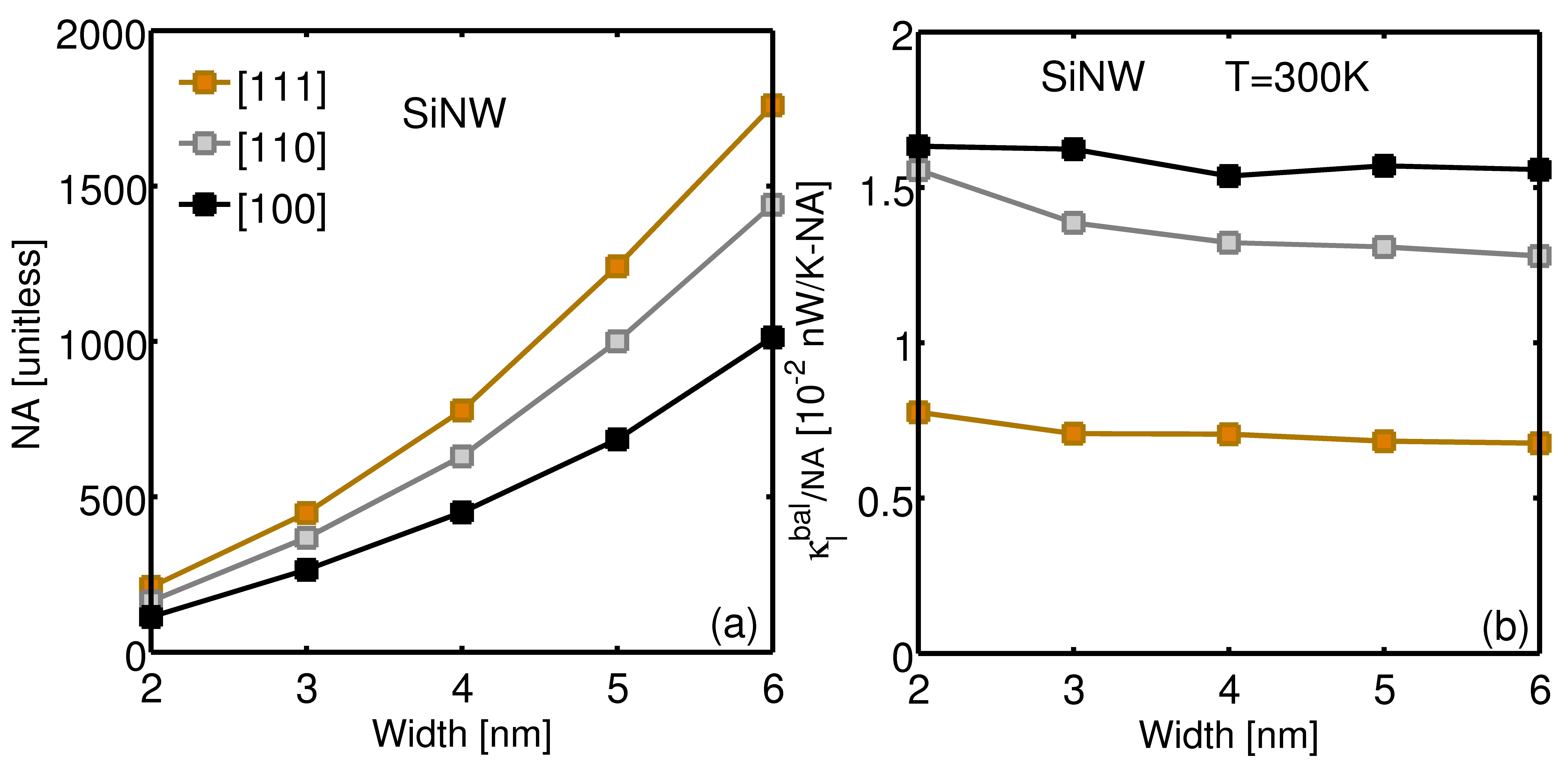}
	\caption{(a) The number of atoms (NA) with W in one SiNW unit cell for different orientations. (b) Contribution to $\kappa_{l}$ per atom for different SiNW orientations.}
	\label{fig:kthermal_orient_NA}
\end{figure}

\textit{Influence of wire orientation on $\kappa^{bal}_{l}$:} The thermal conductance is anisotropic in SiNWs with the following order, $\kappa^{110}_{l}\;\;>\kappa^{100}_{l}\;\;>\kappa^{111}_{l}$ (Fig. \ref{fig:kthermal_orient}a). This result is similar to the one reported in Ref. \cite{jauho_method}. The $\kappa_{l}$ value exhibits a linear variation with NA for all the wire orientations (Fig. \ref{fig:kthermal_orient}b). The width parameters for the thermal conductance (Eq. \ref{eq_kbal_w} ) for different wire orientations are provided in Table \ref{table_3_1}. The order of the ballistic thermal conductance with W for different orientations can be understood by the product of NA$\times m_{k}$ ($P_{nm}$). The NA has the following variation, $NA^{111}\;\;>NA^{110}\;\;NA^{100}$ (Fig. \ref{fig:kthermal_orient_NA}a), whereas $m_{k}$ depicts the following order $m^{100}_{k}>\;\;>m^{110}_{k}\;\;>m^{111}_{k}$ (Fig.\ref{fig:kthermal_orient_NA}b). These two orders are opposite to each other. However, the product shows the following order, $P^{110}_{nm}\;\;>P^{100}_{nm}\;\;>P^{111}_{nm}$). Thus, [110] wires give the highest $\kappa_{l}$ due to the optimal value of NA and $m_{k}$.

An important point to note is that $\kappa^{bal}_{l}$ is expected to decrease further in smaller wires due to phonon scattering
by other phonons, interfaces and boundaries \cite{mingo_ph,Mingo_kappa} which are neglected in this present study. The main idea here is to understand the geometrical effects on the phonon dispersion and the lattice thermal properties of these small nanowires which is attributed to (i) the modification of the phonon dispersion, and (ii) phonon confinement effects in the coherent phonon transport regime.

\subsection{Discussion}
\label{sec_discuss}

In this work, all the thermal properties are shown to scale with W. Also NA depends on W as follows,

\begin{equation}
\label{NA_W_relation}
	 NA  \propto   W^{\gamma},
\end{equation}
where $\gamma\;>0$. So using Eq. (\ref{NA_W_relation}), (\ref{eq_cv_w}) and (\ref{eq_kbal_w}) can be recasted in terms of NA as, 

\begin{eqnarray}
\label{CV_NA_relation}
	\Delta C_{v} (NA) & = & C_{0}\cdot (NA)^{\frac{-1}{\gamma}}  \\ \nonumber
		   & = & C_{0}\cdot (NA)^{-\eta}  \\
\label{kbal_NA_relation}
	 \kappa^{bal}_{l}(NA) & = & K_{0} \cdot (NA)^{\frac{d}{\gamma}}  \\ \nonumber
	 		      & = & K_{0} \cdot (NA)^{\rho}, 
\end{eqnarray}
where, $C_{0}$ and $K_{0}$ are the pre-factors. Thus, a universal power law can be derived for the thermal properties depending on the number of atoms per unit cell (NA) which represents the atomistic effect on the thermal quantities. In these SiNWs, 1.98 $\le\;\gamma\;\le$ 2.1 which provides the limits for $\rho$ and $\eta$,

\begin{eqnarray}
\label{eq_rho}
	 \rho  & \in & [0.93\;,\;1.03]  \\
\label{eq_eta}
	  \eta & \in & [0.48\;,\;0.50]  
\end{eqnarray}

\begin{figure}[htb!]
	\centering
		\includegraphics[width=2.3in,height=4.2in]{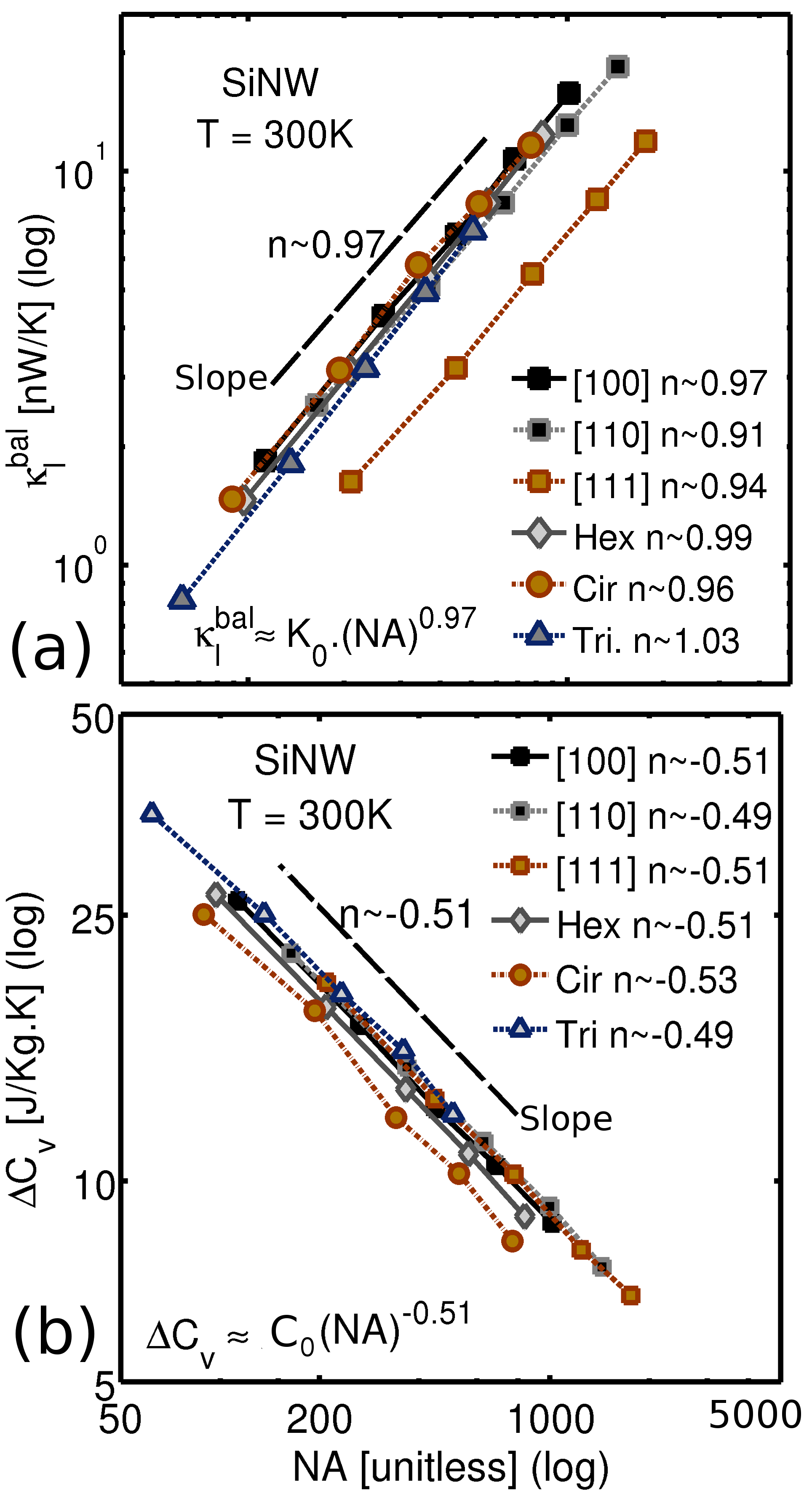}
	\caption{(a) The number of atoms (NA) in one [100] SiNW unit cell for different cross-section size and shapes. (b) Contribution to $\kappa_{l}$ per atom for different cross-section shapes.}
	\label{fig:thermal_NA_dep}
\end{figure}

The variation in the thermal conductance with NA is illustrated on a log-log scale in Fig. \ref{fig:thermal_NA_dep}a. All the SiNWs depict almost the same power law with an average exponent value of $\sim$0.97, which is in the limit derived in Eq. \ref{eq_rho}. Similarly the variation in $C_{v}$ with NA is plotted on a log-log scale in Fig. \ref{fig:thermal_NA_dep}b. All the SiNWs used in this study show the same power law (Eq. \ref{CV_NA_relation}) with an average exponent of -0.51, which is in the limit derived in Eq. \ref{eq_eta}. Thus, the thermal quantities show a universal power law behavior with the number of atoms in the unit cell (NA) irrespective of the details of the unit cell. The details of shape and orientation are embedded in the pre-factors $C_{0}$ and $K_{0}$ (Eq. \ref{CV_NA_relation} and \ref{kbal_NA_relation}).

The relationship of $C_{v}$ and $\kappa_{l}$ to the shape, size and orientation of SiNWs have been provided explicitly in Eqs. (\ref{eq_cv_w}), (\ref{eq_cv_svr}), (\ref{eq_kbal_w}), and (\ref{eq_kbal_na}). These closed form analytical expressions are very handy for the compact modeling of the thermal and thermoelectric properties of SiNWs \cite{Hot_spice,Tspice_tool}. Since these expression are derived from physics-based model, they capture the important geometrical and atomistic effects, thus enabling fast modeling of realistic systems. 
%This proves the utility of the present work from a thermal compact modeling point of view.

\section{Conclusions}
\label{sec:conc}

We have shown the application of the MVFF model for the calculation of the thermal properties of SiNWs. It has been shown that at the nanometer scale these thermal properties are quite sensitive to the wire cross-section size, shape, and orientation. Analytical expressions for the size dependence of the thermal properties of SiNWs of different cross-section shape and channel orientation have been provided. They can be used as component for the compact modeling of the thermal properties of ultra-scaled SiNWs. It has been demonstrated that all the SiNWs follow a universal power law for the specific heat and the thermal conductance which reveals the impact of the atomistic details on these properties. The triangular SiNWs show a high $C_{v}$ and low $\kappa_{l}$, thus making them good candidates for thermoelectric devices. The [110] oriented square Si nanowires are better in terms of heat dissipation due to their high thermal conductance and are therefore good candidates for transistors from a heat management point of view. 

%Cross-section shape along with size and channel oreintation provides another degree of freedom to tune both the electronic and thermal properties in ultra-scaled nanowires like ZT tuning in Ge nanowires \cite{shape_dep_ZT_GeNW} 

\section*{Acknowledgments}

The authors acknowledge financial support from MSD Focus Center, under the Focus Center Research Program (FCRP), a Semiconductor Research Corporation (SRC) entity, Nanoelectronics Research Initiative (NRI) through the Midwest Institute for Nanoelectronics Discovery (MIND), NSF (Grant No. OCI-0749140) and Purdue University. Computational support from nanoHUB.org, an NCN operated and NSF (Grant No. EEC-0228390) funded project is also gratefully acknowledged.

%\nocite{*}
%\bibliographystyle{natbib}
%\bibliography{refs}
%Merlin.mbs v4.21 2009-07-09.
%

\newpage%%
\begin{center}
\textbf{TABLES}
\end{center}

\begin{table}[htb!]
\centering
\caption{Width Parameter for $C_{v}$ in $[$100$]$ SiNWs at T = 300K.}
\label{table_2}
\begin{tabular}{lcc}\hline\hline
Shape & $C^{bulk}_{v}$  & A\\
 units $\rightarrow$ & $(J/kg\cdot K)$  & $(J\cdot nm/kg \cdot K)$\\
\hline\hline
Circular & 681.4 & 47.82 \\
Hexagon & 681.2 & 53.58 \\
Square & 680.9 & 52.66 \\
Triangular & 681.0 & 73.43  \\
\hline\hline
\end{tabular}
\end{table}

\begin{table}[b!]
\centering
\caption{Width Parameter for $C_{v}$ in SiNWs at T = 300K.}
\label{table_2_1}
\begin{tabular}{ccc}\hline\hline
Orientation & $C^{bulk}_{v}$  & A\\
 units $\rightarrow$ & $(J/kg\cdot K)$  & $(J\cdot nm/kg \cdot K)$\\
\hline\hline
$[$100$]$ & 680.9 & 52.66 \\
$[$110$]$ & 681.5 & 43.03 \\
$[$111$]$ & 681.2 & 39.4 \\
\hline\hline
\end{tabular}
\end{table}

\begin{table}[b!]
\centering
\caption{Width Parameter for $\kappa^{bal}_{l}$ in $[$100$]$ SiNWs at T = 300K.}
\label{table_3}
\begin{tabular}{lcc}\hline\hline
Shape & $\kappa_{0}\;\;(nW/K)$  & d \\
\hline\hline
Circular & 0.133 & 1.92 \\
Square & 0.141 & 1.96\\
Triangular & 0.062 & 1.97  \\
Hexagon & 0.097 & 2.01 \\
\hline\hline
\end{tabular}
\end{table}

\begin{table}[!b]
\centering
\caption{Width Parameter for $\kappa^{bal}_{l}$ in SiNWs at T = 300K.}
\label{table_3_1}
\begin{tabular}{ccc}\hline\hline
Orientation & $\kappa_{0}\;\;(nW/K)$  & d \\
\hline\hline
$[$100$]$ & 0.141 & 1.96 \\
$[$110$]$ & 0.204 & 1.87\\
$[$111$]$ & 0.129 & 1.88  \\
\hline\hline
\end{tabular}
\end{table}

\newpage%%
\begin{center}
\textbf{FIGURES}
\end{center}

\end{document}